\documentclass[12pt]{article}
\usepackage{mathrsfs}
\usepackage{graphicx}
\usepackage{amssymb,amsmath,amsthm}
\usepackage{epstopdf}
\usepackage{tocbibind}
 \textwidth = 6.5 in \textheight = 9 in \oddsidemargin =
0.0 in \evensidemargin = 0.0 in \topmargin = 0.0 in \headheight =
0.0 in \headsep = 0.0 in
\def\ba{\begin{eqnarray}}
\def\ea{\end{eqnarray}}

\def\rs{r_s}
\def\del{\delta r}

\def\:{\boldsymbol{:}}

\begin{document}
\begin{flushright}
\end{flushright}
\begin{center} 
\vglue .06in
{\Large \bf {Galactic Halos and Black Holes in Non-Canonical Scalar Field Theories}}\\[.5in]

{\bf Ratindranath Akhoury and Christopher S. Gauthier}\\
[.1in]
{\it{Michigan Center for Theoretical Physics\\
Randall Laboratory of Physics\\
University of Michigan\\
Ann Arbor, Michigan 48109-1120, USA}}\\[.2in]
\end{center}
\begin{abstract}
\begin{quotation}
We consider static spherically symmetric solutions of a general scalar field theory with non-standard kinetic energy coupled to gravity with a view to explain dark matter halos as a coherent state of the scalar field. Consistent solutions are found with a smooth scalar profile which can describe observed rotation curves. A characteristic feature is that the rotation curves are linear for small $r$. The nontrivial solutions have negative scalar energy density near the origin though the total energy is positive definite. We also reconsider the no scalar hair theorems for black holes  with emphasis on asymptotic boundary conditions and super-luminal propagation. Some modifications and extensions of previous analyses are discussed.
\end{quotation}
\end{abstract}
\newpage
\section{Introduction} Recently there has been a lot of interest in the applications of 
 non-canonical scalar field theories in cosmology. This is due mainly to the fact that the energy momentum tensor in such field theories has the potential to describe cosmological fluids with negative pressure which is a necessary ingredient for accelerated expansion. Examples are k-essence \cite{k-essence} which attempts to explain the accelerated expansion of the universe as well as the coincidence problem, the DBI action \cite{dbi}, tachyonic matter \cite{tachyon,ag}, the ghost condensate model \cite{ghost} and the Chaplygin gas model \cite{cgas}. They are also interesting in the context of inflationary models \cite{kinflation,ag2}. Since these theories contain non-standard kinetic energy terms, the constraints of global hyperbolicity and absence of super-luminal propagation play an important role \cite{constraints,armendizlin,vikman}.

In this paper we consider static, spherically symmetric solutions of the Einstein equations coupled to a general k-essence scalar field. The physical motivation is to look for consistent solutions describing galactic halos \cite{salucci} and black holes. The standard assumption concerning galactic dark matter halos is that they consist of an incoherent collection of weakly interacting massive particles. There have been attempts \cite{halos,armendizlin} where some very special type of scalar theories were used to discuss the possibility that the galactic halos could in fact be considered a coherent excitation of a scalar field, much like a boson star. This scenario would have the advantage of  providing a unified treatment of dark matter and dark energy since the latter can be described by a k-essence like theory to begin with. 

We consider the most general scalar field lagrangian which depends on the field $\phi$ and invariants 
of the kind $X = g^{\mu\nu}\partial_{\mu}\phi\partial_{\nu}\phi$ formed from its first derivatives. We do not assume that the kinetic terms are separable nor that they are a quadratic form in the first derivatives.
We find that solutions do exist which can describe galactic halos and for certain choices of the metric function give a good description of the observed rotation curves. Two classes of solutions are found: those that have negative energy density near the origin and those that don't. These can be phenomenologically distinguished by the shape of the rotation curves near the origin. Strictly speaking, only one of these solutions has a valid classical description, whereas the other may depend on new physics at short distance scales. For the case when the scalar energy density becomes negative near the origin, the total energy or mass enclosed can still be positive. Recent evidence from string theory indicates that there is no justification for restricting to potentials that are positive everywhere as long as the total energy is finite \cite{hertog}. Negative energy densities are not unphysical by themselves as the Scharnhorst effect \cite{scharnhorst} clearly shows. Here one has faster than light propagation in a Casimir vacuum. In the literature there are  studies \cite{ftl} associating negative energy densities with super-luminal propagation of signals. However, in \cite{vikman} it was argued that super-luminal propagation does not necessarily imply the violation of causality for which one requires closed time-like curves. All this clearly deserves further study to see if super-luminal propagation without causality violation is consistent. If so, then k-essence like theories would provide a larger context in which to study the formation of galactic halos.

Another question discussed in this paper is the existence of black hole solutions in the combined gravity, k-essence system.  This problem is of physical interest since galaxies are thought to have black holes at their center. Hence, we would like to ascertain whether black holes can coexist with the halos. Two possibilities are usually mentioned in connection with the issue of black hole-halo coexistence \cite{bean}. The first is that the massive black holes were formed together with the galaxies through some internal dynamical process, or secondly that the black holes are primordial. In the second possibility, they were present even before any luminous activity, and in fact are the source driving the quasars. In both of these situations, once formed, the black holes continue to grow in time. If there is a scalar field pervading the universe then it can interact with the black holes and if the scalar no hair theorems \cite{bekenstein} for static spherically symmetric black holes are valid it could cause their accretion. One could study this question by considering the stability of the halo in the presence of a background black hole solution. In this paper, however, we do not do this. Instead our more modest approach is to revisit the scalar no hair theorems. These are sometimes used  to argue \cite{bean} that the black holes become heavier in time by "eating" the scalar hair. We present new ways to understand the theorems and suggest some avenues how they may be circumvented in the context of k-essence like theories.

The paper is organized as follows. In section 2 we discuss the various steps that are necessary to model a spherically symmetric scalar halo and to describe the rotation curves. In section 3 we address the question of the no hair theorems for black holes. We conclude in section 4 with a discussion of our results.

\section{Scalar Fields and Dark Matter Halos} In this section we will describe a halo assuming that it is made up only of scalar fields. Including an exponential disc of baryonic matter should not change the essential conclusions. Thus, we are interested in a scalar field theory coupled to gravity and described by an action of the general form,
\begin{eqnarray}
S = \int d^4x \sqrt {-g}L(X, \phi), \\
X = g^{\mu\nu}\nabla_{\mu}\phi\nabla_{\nu}\phi.
\end{eqnarray}
Its energy momentum tensor is given by,
\begin{equation}
T_{\mu\nu} = {2 \over \sqrt {-g}} {\delta S \over \delta g^{\mu\nu}} = 2L_X \nabla_{\mu}\phi\nabla_{\nu}\phi- g_{\mu\nu}L.
\end{equation}
Here and elsewhere in this paper, $L_X$ denotes the partial derivative with respect to $X$. The equation of motion for the scalar field is given by,
\begin{eqnarray}
\bar{G}^{\mu\nu}\nabla_{\mu}\nabla_{\nu}\phi = {1 \over 2}(L_{\phi} - 2XL_{X\phi}) \label{scalareqn} \\
\bar{G}^{\mu\nu} = L_Xg^{\mu\nu} + 2L_{XX}\nabla^{\mu}\phi\nabla^{\nu}\phi.
\end{eqnarray}
The quantities satisfy several constraints which we now outline: an examination of the characteristics of the scalar equation of motion gives the speed of propagation of the scalar fluctuations, and demanding that this so called sound speed ($C_s$) is not super-luminal imposes the constraint ${L_{XX} \over L_X} \geq 0$. Demanding in addition that the initial value problem be well posed, and the scalar field equation of motion be globally hyperbolic gives the following list of constraints for this system \cite{constraints,armendizlin,vikman},
\begin{equation}
L_X > 0,~~ L_{XX} > 0,~~ {L_X + 2XL_{XX}} > 0.
\end{equation}
In addition, stability requires \cite{vikman} that $C_s > 0$.

As discussed in the introduction we will be interested in static spherically symmetric solutions of the combined Einstein-scalar system of equations. These solutions must match on to the cosmological solution at large $r$. Depending on the model under consideration, the cosmological solution could  either be almost asymptotically flat or not. We will mostly assume almost asymptotically flat solutions, but in the next section we will also consider other possibilities as well.
The space-time metric will be described by,
\begin{equation}
ds^2 = e^{\nu}dt^2 - e^{\lambda}dr^2 -r^2(d\theta^2 + sin^2\theta d\phi^2),
\end{equation}
where $\nu$ and $\lambda$ are functions of $r$ alone. The $(tt), ~(rr)$ and $(\theta\theta)$ components of the Einstein equations $G^{\mu}_{\nu} = \kappa T^{\mu}_{\nu}$ are given by:
\begin{eqnarray}
{e^{-\lambda} \over r^2}(-1 + e^{\lambda} + r\lambda') = -\kappa L, \label{einsteintt} \\
{e^{-\lambda} \over r^2}(-1 + e^{\lambda} - r\nu') = -\kappa (L - 2XL_X), \label{einsteinrr} \\
- {e^{-\lambda} \over 2r}(r\nu^{''} + {1 \over 2}r(\nu')^2 + \nu' - \lambda' - {1 \over 2}r\nu'\lambda') = -\kappa L. \label{einsteintheta}
\end{eqnarray}
The equation of motion for the scalar field (\ref{scalareqn}) in the static limit may be written as,
\begin{equation}
\psi'' + [{\lambda'  \over 2}+ ({\nu' \over 2} + {2 \over r})({L_X \over L_X + 2XL_{XX}})]\psi'
+ {1 \over 2}\{{L_\phi - 2XL_{X\phi} \over L_X + 2XL_{XX}}\} = 0, \label{scalareqn1}
\end{equation}
where $\psi' = e^{-\lambda}\phi'$, and the speed of sound $C_s$ is given by,
\begin{equation}
\delta(r) = {1 \over C_s^2} = {L_X \over L_X + 2XL_{XX}}.
\end{equation}
Note that if we require the absence of super-luminal propagation of the scalar field fluctuations then for all $r$, $\delta(r) \geq 1$, and stability necessitates $C_s > 0$ as well.

We are now ready to discuss the question of galactic halos.  In ref \cite{lake}, it is shown how some essential features of the metric function $\nu$ can be deduced  directly from the observed galactic rotation curves independent of the matter content and a specific gravitation lagrangian.  From stability
considerations \cite{lake} concludes that,
\begin{equation}
 0 < r\nu'/2 <1.  \label{inequality1}
 \end{equation}
Further assuming circular halos and that information travels to us along null geodesics, one get's an additional constraint \cite{lake}, which for our purposes may best be written as:
\begin{eqnarray}
|v'_c(r)| < {v_c(1-v_c^2) \over r}, \\
{r\nu' \over 2} = v_c^2.
\end{eqnarray}
In the above, $v_c$ is essentially the tangential component of the velocity \cite{mtw}, i.e., the rotation curve. These inequalities indicate a smooth scalar profile.
It is clear from these that it is not at all difficult to find functions $v_c(r)$ which give a realistic representation of the rotation curves: one that grows for small $r$ and flattens out in an intermediate region before matching on to the cosmological solution outside. This is precisely the kind of behavior we demand of the corresponding derivative of the metric function $\nu$ so that our model is able to describe the observed rotation curves.

Let us first look for consistent solutions of the different equations of motion near the origin. First, consider the scalar field equation (\ref{scalareqn1}). Since $e^{-\lambda} \neq 0$ near the origin, we rewrite it in terms of the variable $\eta' = e^{-{\lambda \over 2}}\phi'$.  The advantage is that now the scalar equation may be written without explicit reference to the metric function $\lambda$:
\begin{eqnarray}
\eta'' + {1 \over r}({r\nu' \over 2} + 2)\delta(r)\eta' + h(r)= 0, \label{scalareqn2} \\
\textrm{where}\quad h(r) = {L_{\eta} - 2XL_{X\eta} \over 2(L_X + 2XL_{XX})}. 
\end{eqnarray}
We would like to determine the appropriate boundary conditions for small $r$. Since we are describing a halo as a coherent state of a scalar field, much like a boson star, the appropriate boundary condition at the origin is the generalized ``no force'' condition, i.e. ${dp \over dr} = 0$, where $p$ is the pressure
and ${dp \over dr}$ is given by the Oppenheimer-Volkov equation.
For our model, the pressure and the density are respectively, $\rho = - L$ and $p = L - 2XL_X$.
Using (\ref{scalareqn1}) or (\ref{scalareqn2}), it is straightforward to obtain:
\begin{equation}
{dp \over dr} = - {2 \over r}({r\nu' \over 2} + 2)\eta'^2L_X. \label{dpdr}
\end{equation}
Since the quantity inside the brackets is bounded between $2$ and $3$, we obtain the boundary condition for the scalar field equation that $\eta'$ goes to zero faster than $\sqrt r$ at the origin (keep in mind that $L_{X} >0$ everywhere). We would next like to determine the exact rate at which this happens, so for this purpose  consider a small $r$ expansion for $\eta$  and write,
\begin{equation}
\eta = r^s(c_0 + c_1r + c_2r^2 + \cdots). \label{expansion}
\end{equation}
As a consequence of the condition that $\eta'$ vanish at the origin, it follows that $h(r)$ must have a smooth behavior at small $r$. To see why this is the case, we will first assume the opposite, i.e. suppose that $h(r)$ goes like ${1 \over r}$ to leading order at small $r$. The equation of motion (\ref{scalareqn2}) implies that to leading order $\eta'$ approaches a nonzero constant at small $r$, which rules out this possibility. Similarly, if $h \sim 1 / r^{n}$, where $n \geq 2$ is an integer, then $\eta' \sim 1/r^{n-1}$, which would contradict our previous finding that $\eta' \rightarrow 0$ at the origin. So the only allowed behavior of $h(r)$ consistent with the boundary conditions is that it is well defined at the origin. Substituting (\ref{expansion}) into (\ref{scalareqn2}), we find for the indicial equation ($c_0 \neq 0$):
\begin{eqnarray}
s(s - 1 +\bar{\gamma} \bar{\delta}) = 0 \\
2 < \bar{\gamma} < 3.
\end{eqnarray}
In the above, $\bar{\gamma}$ denotes the value of $\gamma = r\nu'/2 +2$ for small $r$ and the inequality  above follows from (\ref{inequality1}). Further, if we demand the absence of super-luminal propagation of the scalar fluctuations then $\bar{\delta}$, the value of $\delta$ for small $r$, is greater than 1. Thus we see that the absence of super-luminal propagation of the scalar fluctuations forces upon us the solution with $s=0$. The actual behavior of $\eta'$ with $r$ near the origin depends on the small $r$ behavior of $h(r)$\footnote{To consider an extreme case, if the lagrangian depends only on $X$ then this term is absent. It is then trivial to verify that the only solution which is smooth at the origin is $\eta = $constant. This in turn implies the constancy of the pressure and the energy density in this case. This is the de Sitter solution. We will not discuss this possiblity further.}. By substituting the expansion (\ref{expansion}) with $s=0$ into (\ref{scalareqn2}) we find $c_1 =0$ if $h(r)$ is well behaved near the origin, and it is non-zero when $h(r) \sim 1/r$. Thus depending on the small $r$ behavior of $h(r)$, $\eta'$ either vanishes at the origin or approaches a constant. In either case since $c_0 \neq 0$, $\eta$ approaches a constant at small $r$. The boundary condition for our case is that $\eta' \rightarrow 0$ at small $r$, thus the only behavior of $h(r)$ consistent with the absence of super-luminal propagation of the scalar fluctuations is that it is either a constant or vanishing at small $r$.  
In either case, $c_1 = 0$, which is what we henceforth take to be the case. We will see below how these conditions translate into the shape of the rotation curves for small $r$. 

We must now find a consistent solution for small $r$ for the metric functions as well. For these we
turn to the Einstein equations (\ref{einsteintt}) and (\ref{einsteinrr}). It is useful to rewrite these as
\begin{eqnarray}
\lambda' = -\kappa rL - (e^{\lambda} - 1)(\kappa rL + {1 \over r}), \label{lambdap} \\
\nu' = \kappa rp + (e^{\lambda} - 1)(\kappa rp + {1 \over r}). \label{nup}
\end{eqnarray}
On the right hand side of these equations we have the energy density $\rho = -L$ and the pressure $p=L-2XL_X$. From these we see that the combination $(\nu' + \lambda')$ only depends on $p-L$, which is proportional to $X$ (see also (\ref{einsteincombine}) below). $\nu$ and $\lambda$ themselves depend  on $p$ and $L$ separately. In general, the small $r$ behaviors of $L$ and $(p-L) \sim X$ can be different: $L$ is a function of  both $X$ and $\phi$, and $\phi$ goes to a constant at small $r$. From a Taylor expansion of $L$ around $X=0$\footnote{We assume that the lagrangian is Taylor expandable in $X$.}, it is seen that the leading behavior of $L$ can therefore be either a constant or that of at least $X$.
From (\ref{lambdap}) and (\ref{nup}) we see that in the former case, the leading behavior of $\nu'$ and 
$\lambda'$ are determined from that of $L$ and this leading behavior is cancelled from the sum. It is the subleading terms of $\nu'$ and $\lambda'$ which are now proportional to $X$. 
For small $r$, we have two possibilities: (i) both $L$ and $2XL_X$ have similar behavior near the origin, i.e. since $L_X > 0$, $L \sim 2XL_X \sim r^2$  at most or, (ii) $L \sim$ constant, and $2XL_X \sim r^2$ or faster. In the case of possibility (i) the scalar field and in particular the speed of sound plays an important role in restricting the small $r$ behavior. In contrast, for possibility (ii) the leading behavior is governed by $L$ and the constraints on the absence of super-luminal propagation of the scalar fluctuations do not play a role in determining the small $r$ behavior of the metric functions. We will discuss each possibility below.

For possibility (i), it is now straightforward to check that consistency with the Einstein equations gives the following behaviors at small $r$ for the metric functions: $\nu, \lambda \sim r^4$ or faster. At this point the scalar energy density and pressure go like $r^2$ or faster. This behavior excludes a purely $X$ independent potential term in the lagrangian for the scalar field since that could lead to a constant behavior for small $r$ for $\rho = - L$. The sign of the energy density will now be considered.
Equation (\ref{dpdr}) is useful for this purpose as well. The term in brackets on the right hand side of this equation is positive so $p$ is monotonically decreasing as we go from some small $r$ to infinity where it approaches the (almost zero) value dictated by the near asymptotically flat cosmological solution outside the galactic halo. The pressure at small $r$ is therefore positive. Since $p$ is positive and monotonically decreasing to zero for all $r$, one should conclude that because in case (i), $p = L - 2 X L_{X} \sim r^{2}$, the only possibility is the trivial solution $p=0$ everywhere. While this would be true if the classical theory is valid everywhere, it is possible that quantum gravity effects modify the theory at small scales negating this argument. The influence of the constraint of subluminal propagation of the scalar fluctuations has in fact a very indirect influence on all this. The real reason for the absence of the negative energy density is that demanding the leading small $r$ behavior of $\eta'$ to be relevant both for the pressure and the lagrangian, eliminates the pure potential terms from consideration. We will discuss this in more detail below.

Possibility (ii) leads to a completely different conclusion. In this case it is easy to check that the metric functions have the following behavior for small $r$: $\nu, \lambda \sim r^2$. The essential difference with the first possibility is that since eq. (\ref{dpdr}) implies that the maximum positive value of $p$ is for small $r$, therefore, $p \approx L > 0$. Thus the energy density $\rho = -L$ can now be negative for small $r$. The lagrangian can contain terms which depend only upon $\phi$ and not its derivatives. Such potential terms can approach a (negative) constant at small $r$. This case is analogous to the flat space situation discussed in \cite{ag}. As we will see below, this should not be surprising since the condition for the absence of super-luminal propagation of the scalar fluctuations has not played a role here. Even though the energy density can be negative in some region, we will now see that under very reasonable assumptions  the total mass within a large enough region is positive definite. The metric function $\lambda$ is related to the mass function $m(r)$ by 
\begin{equation}
e^{\lambda} = (1 - {2Gm \over r})^{-1}. \label{massdef}
\end{equation}
Consider the total mass inside a radius $R$. From (\ref{einsteintt}) and (\ref{massdef}) we have,
\begin{equation}
m = -4\pi\int_0^R L r^2dr = -4\pi\int_0^R (p + 2XL_X)r^2dr. \label{mass}
\end{equation}
In the first term in the integral we perform an integration by parts to write,
\begin{equation}
\int_0^R pr^2dr = -{1 \over 3}\int_0^R{dp \over dr}r^3dr + [{1 \over 3}pr^3]_0^R.
\end{equation}
Using (\ref{dpdr}), and substituting into (\ref{mass}), we obtain for the mass parameter $m(R)$,
\begin{equation}
m = {8\pi \over 3}\int_0^R(\eta')^2L_X(1 - {r\nu' \over 2})r^2dr - {4\pi \over 3}[pr^3]_0^R.
\end{equation}
Using (\ref{inequality1}) the first term is manifestly positive definite and the surface term can be made small for large enough $R$ if the pressure falls off faster than ${1 \over R^3}$ (asymptotically near flat condition) or if the cosmological solution is such that for large $r$ the pressure is negative as is the case with some k-essence models.

The origin of the negative energy densities in certain regions can be clearly traced to the purely potential terms in the lagrangian. Indeed, as $\phi$ goes to a constant for small $r$, these potentials tend to a negative constant. If the total mass parameter is positive, then it appears to us that excluding theories where the potential can be negative in some regions is not justified. In fact, potentials such as these 
have been recently considered in a variety of situations. For example, in supersymmetric AdS compactifications one encounters potentials with local negative maximums, and also a large class of supersymmetric compactifications including Calabi-Yau and G2, give rise to effective four dimensional potentials with negative regions \cite{hertog}. The stability of the solutions considered in this paper will be the subject of a future investigation. We would like to note, however, that there are many known examples where the potential is negative at an extremum and yet the solution is stabilized due to gravitational effects \cite{boucher} as long as the scalar field theory satisfies the positive energy theorem \cite{abbot}. 

Another reason to exclude negative energy densities is its association with super-luminal propagation and causality violation and we turn to a discussion of this point. Consider the constraints on super-luminal propagation in some more detail. Until now the only constraint we have considered is the one arising for the speed of the scalar fluctuations, which is obtained from an analysis of the characteristics of the scalar equation of motion. A simple argument shows that this has no direct connection with the sign of the energy density. If one changes $L \rightarrow -L$ the scalar equations of motion does not change and hence neither does the analysis of the characteristics but the energy density changes sign. Therefore there must be an additional constraint on super-luminal propagation which is dependent on the sign of the energy density. In fact, there have     been quite a few papers which investigate the relationship between faster than light travel and negative energy densities \cite{ftl}. They all differ on the precise definition of super-luminal propagation. For the purposes of this paper we follow the discussion of \cite{marolf}, which is specific to static spherically symmetric space-times for which the killing time can be used to measure the time required for objects or signals to propagate between two of its orbits. In \cite{marolf} a theorem was proven that if in such a space-time, the (time-like) weak energy condition is satisfied, then the signaling time is never faster than the corresponding signal in Minkowski space. The normalization of the Killing time is appropriate for an observer at very large distances. More specifically, the absence of super-luminal signals in the sense defined above, requires $T_{\mu\nu}t^{\mu}t^{\nu} \geq 0$ for any time-like vector $t^{\mu}$. Demanding  the absence of super-luminal propagation based on this criterion would eliminate possibility (ii) discussed above since it implies negative energy densities at very small $r$. However, super-luminal travel by itself is not threatening as long as there is no causality violation for which closed time-like curves are required. In fact, recently \cite{vikman} it was realized from the viewpoint of pure classical field theory, that models which allow for super-luminal propagation even on dynamical backgrounds do not necessary posses internal contradictions. In particular these models do not lead to any additional causal paradoxes over and above those already present in standard general relativity.

We will therefore take the point of view that neither of the solutions discussed above can be a priori discarded (if the total mass parameter is positive definite and well behaved) and look for the data to decide if both are realized or not. In this regard it is interesting to note that the two possibilities discussed above can predict very different behavior of the rotation curves for small $r$ depending on the model for the lagrangian: possibility (ii) implies $v_c \sim r$, and, if such a possibility exists in the complete theory, (i) implies $v_c \sim r^2$ or faster. Our discussion suggests that the shape of the rotation curves for small $r$ should therefore provide the phenomenological distinction between the possibilities considered.

If the theories we consider can indeed describe halos of dark matter, then we need to have an understanding of its interactions with a black hole which presumably are at the center of galaxies.
With this in mind, the next section will revisit the "no scalar hair" theorems \cite{bekenstein} for black holes.

\section{Dressing Black Holes With Scalar Fields} In this section we study the possibility of 
 a scalar field dressing an asymptotically flat, static, spherically symmetric black
hole solution in the theory discussed in the earlier sections. As discussed in the introduction, this question is of relevance for the  stability of the halos described previously.

The distinguishing feature of a black hole is the existence of an event horizon whose position
at $r_s$ is determined from the condition, $g^{rr}(\rs)=0$. Physical quantities, in particular the components of the energy momentum tensor, are regular at the horizon, from which we deduce  
the regularity of the scalar field and its derivative at $r=r_s$, instead of regularity at the origin. With this as the boundary condition, let us consider the scalar field equation of motion
near the horizon. We want to find the behavior of the allowed solutions consistent with the absence
of super-luminal propagation of the scalar field fluctuations. From equation (\ref{scalareqn1}) we see 
that we need to know the behavior of $\nu'$ near the horizon. For this purpose let us
consider the $(tt)$ component of the Einstein equation (\ref{einsteintt}) near $r=r_s$ and write, 
\begin{equation}
e^{-\lambda} = A_{\lambda}(r-\rs) + O((r-r_s)^2). \label{horizon1}
\end{equation}
Thus, 
\begin{equation}
-\lambda' = { 1 \over {r-r_s}} + constant + O(r-\rs), \label{lambda'}
\end{equation}
and from (\ref{einsteintt}) we obtain that, 
\begin{equation}
A_{\lambda} = { 1 + \kappa r_s^2L(r_s) \over \rs} > 0. \label{alambda}
\end{equation}
The last inequality follows from the fact that $e^{-\lambda}$ must be growing as we move out from
the horizon. There is the possibility that $A_{\lambda}=0$, which would give the leading behavior in (\ref{horizon1}) as $(r-\rs)^2$. This would correspond to the case of an extremal black hole solution, which we will not discuss further. We are now able determine the leading behavior of 
$\nu$ near the horizon consistent with the Einstein equations (\ref{einsteintt}, \ref{einsteinrr}). From these,
\begin{eqnarray}
(\nu'+\lambda') = e^{\lambda}r\kappa f(r), \label{einsteincombine} \\
f(r) = -2XL_X > 0. \label{f}
\end{eqnarray}
Integrating this in the neighborhood of the horizon we get, 
\begin{equation}
\nu+\lambda = {\rs \over A_{\lambda}} f(\rs)\ln(r-\rs) + K + O(r-\rs),
\end{equation}
where $K$ is an integration constant. Using the previously determined value of $\lambda$, we finally obtain for the leading behavior of $\nu$ assuming $A_{\lambda} \neq 0$,
\begin{eqnarray}
\nu = B\ln(r-\rs) + constant + O(r-\rs), \\
\textrm{where }\quad B = {1 + \kappa\rs^2(L(\rs) + f(\rs)) \over 1 + \kappa\rs^2L(\rs)}.
\end{eqnarray}
It is easy to check from the finiteness of the Ricci scalar at the horizon that $B=1$, which tells us that
$f(\rs)$ approaches zero there. Since $L_X > 0$, we see that $X \rightarrow 0$ as $r \rightarrow r_s$. The exact rate at which this happens will be determined below. From this we see that,
\begin{eqnarray}
\nu' =  { 1 \over {r-r_s}} + constant + O(r-\rs), \label{nu'}     \\
e^{\nu} = A_{\nu}(r-\rs) + O(r-\rs)^2,
\end{eqnarray}
where the constants in (\ref{nu'}) and (\ref{lambda'}) are not related and $A_{\nu} > 0$. 

Let us now look for a series solution of the scalar field which is regular at the horizon. In
addition, we have seen from the discussion immediately preceding Eq. (\ref{nu'}), that the 
solution should be such that $X = -e^{-\lambda}\phi'^2$ must vanish there. Introducing
$\delta r = r-r_s$, we have the following expansion:
\begin{equation}
\psi = {\del}^s\{ a_0 + a_1\del + a_2\del^2 + \cdots\}. \label{expansion2}
\end{equation}
Substituting this into Eq.(\ref{scalareqn1}) we get the following indicial equation,
\begin{equation}
s(2 s+\alpha - 3) = 0,
\end{equation}
where, $\alpha = {1 \over C^2_s(r_s)}$. From the discussion on subluminal propagation for the scalar field fluctuations in the previous section, $\alpha \geq 1$. In this instance, both the solutions for $s$ are not immediately excluded provided $\alpha$ is not greater than 3. The special case of $\alpha = 3$ is the same as the $s=0$ case except the second solution has a leading $\log r$ behavior. Let us first consider the possible solutions with near horizon behavior dictated by $s = 3/2-\alpha/2$. As a result of the smoothness of the action near the horizon, the small $\delta r$ leading order behavior of $\psi'$ is ${\delta r}^{1/2-\alpha/2}$. This behavior is too singular if we are considering only subluminal propagation, and hence, this solution is ruled out. We next consider the solution with $s=0$. For the same reasons discussed in the previous section, $a_1=0$, and the leading near horizon behavior of $\psi'$ is of the form $\delta r$. Moreover, this near horizon behavior implies $\phi'=$ constant or $X \sim e^{-\lambda}\phi'^2 \sim \delta r$. This is completely consistent with the Einstein equation (\ref{einsteincombine}) (Recall that always, $L_X > 0$). Indeed from the combined near horizon behavior  of $\nu'$ and $\lambda'$ we see that the leading term on the LHS of this equation is a constant, which is the same on the RHS. We should comment that the Einstein equation (\ref{einsteincombine}) rules out the possibility $s = 3/2-\alpha/2$ when $\alpha < 1$ (super-luminal propagation). Thus it should be emphasized that consistency (at the horizon) for the $s=0$ solution holds irrespective of whether we have super-luminal or subluminal propagation of the scalar field fluctuations.

We must now check the consistency of the solution at large $r$. The main question we would like to address is whether the solutions at the horizon can match on to the ones at infinity in a manner consistent with the Einstein equations. First let us list the behavior of the pressure $p$ and 
${dp \over dr}$ at the horizon and asymptotically. We will see how far one can go without assuming the weak energy condition.

Consider,
\begin{equation}
{dp \over dr} = - 2({\nu' \over 2} + {2 \over r})\eta'^2L_X, \label{dpdr2}
\end{equation}
near the horizon. Using (\ref{nu'}) and (\ref{horizon1}) and the fact that $\phi' = K$ (a constant), we get,
\begin{equation}
{dp \over dr} \sim - A_{\lambda} K^2L_X < 0.
\end{equation}
In order to find its behavior at infinity we need the asymptotic form of $\nu'$. For this purpose, one could first find the asymptotic form of $\lambda$ from the integrated version of (\ref{einsteintt}):
\begin{eqnarray}
e^{-\lambda} = 1 - {2Gm(r,r_s) \over r} - {r_s \over r}, \\
\textrm{where} \,\,\, m(r,r_s) = 4\pi\int_{r_s}^r \rho r^2dr.
\end{eqnarray}
Assuming that $\rho$ falls off faster than ${1 \over r^3}$ at large $r$, the 
leading behavior of $\lambda'$ is,
\begin{equation}
- \lambda' \sim {2Gm(r,r_s) \over r^2} + {r_s \over r^2}.
\end{equation}
From (\ref{einsteincombine}) and (\ref{f}) it would follow that 
\begin{equation}
{\nu' \over 2} > {(Gm(r,r_s) + {r_s \over 2}) \over r^2} \geq 0,
\end{equation}
which in turn would imply from (\ref{dpdr2}) that ${dp \over dr} < 0$ also asymptotically for large $r$.
However, we want to get information about the large $r$ behavior of ${dp \over dr}$ without using 
the asymptotic flatness condition that $\rho$ falls off faster than ${1 \over r^3}$ at large $r$. For this purpose we note that since ${dp \over dr} < 0$ near the horizon, in order for it to be positive asymptotically  at large $r$, it would have to vanish in between. We will now show that this is untenable on the basis of the Einstein equations, apart from the trivial solution that $\phi$ is a constant. For this purpose note that from (\ref{dpdr2}), when ${dp \over dr} = 0$, 
\begin{equation}
{\nu' \over 2} = - {2 \over r}.
\end{equation}
It is easily checked using (\ref{einsteinrr}) that this implies,
\begin{equation}
e^{\lambda} = - {3 \over 1 + \kappa pr^2}.
\end{equation}
Since we are looking for regular black hole solutions of the combined scalar-Einstein system, we must rule out the possibility of a change in signature of the metric, which means that $1 + \kappa pr^2 < 0$ or
$\kappa pr^2 < -1$. From (\ref{einsteincombine}) and (\ref{f}) however, 
\begin{equation}
- ({2  \over r}){2 + 3\kappa pr^2 \over 1 + \kappa pr^2} > 0,
\end{equation}
or, $\kappa pr^2 > -{2 \over 3}$. The two constraints on $\kappa pr^2$ are incompatible, hence we conclude that ${dp \over dr}$ cannot vanish between the horizon and infinity and must therefore also be negative asymptotically. We emphasize that our conclusion that ${dp \over dr}$ is negative everywhere outside the horizon does not rely on either the weak energy condition or the asymptotic flatness
condition.

Let us next find how the pressure $p(r)$ behaves near the horizon and asymptotically. To find $p$ near the horizon, note that here, ${\eta'}^2 \sim (r-r_s)$. Thus, at the horizon, $p = - \rho$. This is the only place we need to worry about whether $\rho$ is positive or negative. If $\rho$ is negative then $p$ is positive at the horizon and if instead $\rho$ is positive then $p$ is negative there. Let us now consider the possibilities for $p$ at large $r$. For positive $\rho$, since ${dp \over dr}$ cannot change sign in between, $p$ must also be negative at large $r$. Now it is easy to see that if such a solution were to exist, it cannot be asymptotically flat. Indeed, writing $2{\eta'}^2L_X = p - L$, in (\ref{dpdr2}) and integrating using the integrating factor we get,
\begin{equation}
p = -  {e^{-{\nu \over 2}} \over r^2}\int_{r_s}^r(e^{\nu \over 2}r^2)'\rho dr.
\end{equation}
Let us see what this implies for $p(r)$ at asymptotic values of $r$. Using the asymptotic condition that $\rho$ falls off faster than ${1 \over r^3}$ at large $r$, we see that the integral converges and 
$|p|$ falls off at least like ${1 \over r^2}$ at large $r$. However, since we have already argued that
${dp \over dr}$ is negative, it must be that $p$ is positive asymptotically at large $r$. The condition of asymptotic flatness gives a positive value for $p$ at large $r$, thus excluding this possibility. Furthermore, asymptotic flatness is incompatible with the constraint $L_{X}>0$. This is a result of the fact that $L_{X}>0$ implies that $\rho > - p$. Since $p$ is negative in this case and monotonically decreasing, it follows from $\rho > - p$ that $\rho$ has to be positive and monotonically increasing in order for $L_{X} >0$ to be satisfied. The only way we can have an asymptotically flat solution is if $p = 0$ everywhere. We therefore conclude that for positive energy density $\rho$, the only allowed black hole solution has negative pressure asymptotically at large $r$, and does not obey the asymptotic flatness condition. In k-essence models it is possible to have a cosmological solution in the matter dominated epoch with these properties. However, it is not clear how we can match on a dynamically evolving scenario onto a static solution.

When the energy density is negative, the pressure can be both positive or negative for large $r$ indicating a solution where one can get either asymptotic flatness or not. However, if we invoke the results of \cite{marolf} regarding super-luminal propagation of signals discussed in the previous section then this possibility would be excluded. For reasons mentioned earlier, we do not subscribe to this viewpoint as long as the enclosed black hole mass is positive and well behaved. We should mention at this point that solutions which can have negative energy density somewhere should specially be checked for stability. 

Our discussion in this section suggests that the possibility of a black hole and a scalar halo forming at the same time is difficult. This however, does not necessarily make the scalar halo model problematic. It is quite possible, and there are other evidence \cite {bean} in support of  a scenario where a primordial black hole was present before the formation of the galaxy itself, triggering the quasar activity. The halo can then be treated as a perturbation on this primordial black hole background. An analysis of this is beyond the scope of this paper and is the subject of an ongoing project.

\section{Summary and Conclusions} We have found some consistent solutions of the scalar-gravity system, which can describe galactic halos. Solutions which have negative energy density near the origin have rotation curves with $v_c \sim r$ at small $r$. Classically these are the only allowed solutions, however, should solutions that have vanishing pressure at the origin be allowed in a complete theory, the rotation curves should show much steeper behavior for small $r$. Solutions with negative energy density are associated with super-luminal propagation as discussed in section 2, however, that does not necessarily imply causality violation. The total energy of these configurations can be positive definite even if the energy density is negative somewhere. On basis of this, and consistent with similar phenomenon in other physical situations \cite{hertog}, we have argued against excluding such configurations a priori. Our analysis in section 2 is based on spherical halos. However, under reasonable assumptions, we expect similar conclusions in general. This is because we can parametrize the departure from sphericity by a parameter which should have a smooth limit to zero unless there are topological obstructions.

We have also reconsidered the question of  static spherically symmetric black hole solutions in a theory of gravity coupled to scalar fields with non-standard kinetic energy terms. We have considered situations where the scalar energy density can be negative in some regions, but the total mass is still positive as well as the almost asymptotically flat and other boundary conditions. In the case when the (time-like) weak energy condition is satisfied and the asymptotically flat boundary conditions enforced, we recover the scalar no-hair theorems \cite{bekenstein}. Our analysis is based on the minimum of assumptions and different from previous ones. We find loopholes not only for the case of negative energy densities, but also when the boundary conditions are not asymptotically flat. For reasons mentioned earlier, we do not think it is justified to, a priori, discard solutions where the scalar energy density is negative somewhere but where the total black hole mass is positive and well behaved. In this context it should be mentioned that in \cite{vikman2} consistent black hole solutions have been found in the kind of theories considered here which are stationary and do contain regions where the energy density can be negative. We have not discussed the stabilty of any possible solutions. This and the construction of explicit solutions is also an ongoing project.

\section{Acknowledgements} We acknowledge Arjun Menon for some initial collaboration
and would also like to thank him, Fred Adams, Paul Federbush, Saul Teukolsky and Alex Vikman for discussions. We are especially grateful to Ted Jacobson for communications about  super-luminal propagation of the energy flux in general relativity. This work was supported by the US department of energy.

\newpage

\end{document}